\documentclass[english,10pt,aps,prx,twocolumn,superscriptaddress,amsmath,amssymb,floatfix,noeprint,letterpaper]{revtex4-2}
\usepackage{graphicx}
\usepackage{physics}
\usepackage{xcolor}
\usepackage[normalem]{ulem} 
\usepackage{bbm}
\usepackage[colorlinks=true,citecolor=blue,linkcolor=blue,urlcolor=blue]{hyperref}
\usepackage{comment}

\usepackage{times}

\begin{document}

\title{Observing Conformal Floquet Dynamics on a Digital Quantum Processor}

\author{Liang-Hong Mo}
  \altaffiliation{These authors contributed equally to this work.}
  \affiliation{Department of Physics, Princeton University, Princeton, New Jersey, 08544, USA}
  
\author{Bastien Lapierre}
  \altaffiliation{These authors contributed equally to this work.}
  \affiliation{Philippe Meyer Institute, Physics Department, Ecole Normale Supérieure (ENS), Université PSL, 24 rue Lhomond, F-75231 Paris, France}

\author{Qiang Miao}
  \email{Corresponding author, qiang\_miao@outlook.com}
  \affiliation{Duke Quantum Center, Duke University, Durham, North Carolina 27701, USA}

\date{June 2026}

\begin{abstract}

Quantum simulations are traditionally confined to exploring dynamics starting from unentangled or low-entanglement states due to severe bottlenecks in protocol design, hardware performance, and classical verification. Here, we report the first quantum simulation of non-equilibrium dynamics initiated directly from a many-body critical ground state. Using a fully-connected trapped-ion processor, we prepare the critical ground state of a transverse-field Ising model via a hardware-tailored, logarithmic-depth quantum circuit based on multi-scale entanglement renormalization. Following this initialization, we apply a deep Floquet drive that maintains emergent conformal symmetry, enabling us to benchmark lattice dynamics against analytical results from continuum theory. In the resulting conformal heating phase, we extract a central charge consistent with the Ising universality class ($c=1/2$) from the universal decay of the Loschmidt echo and observe spatial energy localization predicted by field theory. Conversely, the non-heating phase exhibits global finite-time revivals. This work establishes a scalable and versatile framework for exploring critical quantum dynamics.

\end{abstract}

\maketitle

\section{Introduction}\label{sec:intro}

Programmable quantum processors offer a native, scalable route for simulating quantum many-body dynamics~\cite{Feynman1982-21,Georgescu2014-86,bernien2017probing,zhang2017observation, daley2022practical}, yet limited coherence times demand efficient circuit design for state preparation. Classical tensor network methods, meanwhile, have proven remarkably powerful for representing well-structured many-body states in equilibrium~\cite{White1992-69,White1993-48,Schollwock2011-326}, but face a computational bottleneck when entanglement grows under non-equilibrium dynamics~\cite{schuch2007entropy,osborne2006efficient}.
It is then natural to ask whether combining these two approaches can preserve the merits of each, opening a promising route to simulating dynamics starting from complex many-body states, which is hard to access by either approach alone.

Within the landscape of quantum many-body states, critical states occupy a central place, as their low-energy physics is universal and insensitive to microscopic details.
In one spatial dimension (1D), this universality is described by conformal field theory (CFT)~\cite{cardy1996scaling, DiFrancesco1997}, which captures long-distance correlations and entanglement entropy. 
Recent quantum simulation experiments have successfully probed such CFT signatures in critical ground states, including power-law correlations, entanglement entropy scaling, and emergent low-energy spectrum~\cite{Haghshenas_2024, 2025sciencenormanyao, miao2025probingentanglementscalingquantum, Koyluoglu2026,sun2026}. 
Beyond ground state physics, CFT has offered numerous insights into universal quench dynamics and thermalization in critical systems~\cite{Calabrese_2005}, and more recently into heating dynamics in driven systems~\cite{wen2018_05, Fan2020-10,PhysRevResearch.2.023085, Wen_2021_PRR,Lapierre_2021,dvlw-hl7t,sg51-1c1s}. Realizing this non-equilibrium universality requires that the time evolution remains within the low-energy sector, a condition that product-state initializations cannot satisfy: they inject high-energy excitations at finite density into a gapless spectrum, which obscures signatures of universality and generically leads to featureless infinite-temperature heating~\cite{PhysRevX.4.041048, PhysRevE.90.012110}.

In this work, we report the \textit{first} quantum simulation of non-equilibrium dynamics starting from a many-body critical ground state. 
Using Quantinuum's trapped-ion digital quantum processor, we prepare the ground state of a critical Ising model via a hardware-tailored, logarithmic-depth quantum circuit based on the multi-scale entanglement renormalization ansatz (MERA)~\cite{Vidal2007-99,Vidal2008-101}.
Starting from this critical state, we then apply a Floquet circuit to explore universal driven dynamics, finding quantitative agreement with emergent CFT descriptions.
By tracking the time evolution of two complementary observables: the Loschmidt echo and the energy density, we observe robust signatures of the heating and non-heating dynamical phases sustained over 16 Floquet cycles.

Furthermore, we propose and realize a protocol to extract the central charge $c$ of the underlying critical model from the time evolution of the Loschmidt echo. 
We derive a universal closed-form expression for the Loschmidt echo of the evolved critical ground state (see App.~\ref{sec:cft_prediction}), showing that its evolution is fully determined by conformal invariance, with the central charge as the \textit{only} theory-dependent parameter. 
Fitting this single parameter to the measured Loschmidt echo establishes a new route to probe the central charge in critical models, bypassing the heavy measurement overhead of entanglement-based observables.
The proof-of-principle demonstration here is not only grounded in low-energy universality but also scalable. Our work thus establishes a generic framework for realizing non-equilibrium dynamics from a scale-invariant vacuum rather than a product state. 

\begin{figure*}
    \centering
    \includegraphics[width=0.95\linewidth]{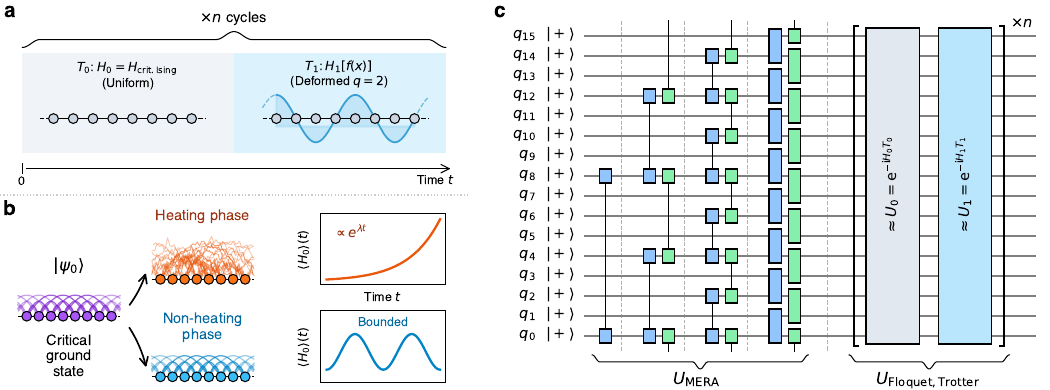}
    \caption{\textbf{Setup and implementation of Floquet circuit.} (a) We consider a periodic driving protocol alternating between the uniform TFIM Hamiltonian $H_0$ and the deformed TFIM Hamiltonian $H_1$, with initial state prepared in the ground state of $H_0$ . In each cycle, the system evolves under either $H_0$ or $H_1$ for durations $T_0$ and $T_1$, respectively. (b) Depending on driving parameters, the system exhibits either a heating phase where energy grows exponentially, or a non-heating phase where energy remains bounded. (c) Circuit implementation of the ground state preparation and time-evolution protocol. The target critical ground state is generated via the hierarchical $U_{\mathrm{MERA}}$ circuit, where each two-qubit gate is parameterized as $\mathrm{e}^{-\mathrm{i}(\phi Y \otimes Z + \gamma Z \otimes Y)/2}$. Following state preparation, the system is subjected to a nonequilibrium Floquet drive for up to $n=16$ cycles. The alternating time-evolution operators, $U_0 = \mathrm{e}^{-\mathrm{i} H_0 T_0}$ and $U_1 = \mathrm{e}^{-\mathrm{i} H_1 T_1}$, are approximately implemented via a second-order Trotter-Suzuki decomposition~\eqref{eq:trotter}.}
    \label{fig:setup}
\end{figure*}

\section{Conformal Floquet dynamics}\label{sec:cfd}

Our goal is to digitally simulate the driven dynamics of a 1D critical spin chain and to extract universal signatures of the underlying conformal symmetry. We consider a transverse-field Ising model (TFIM) with site-dependent couplings and periodic boundary conditions,
\begin{align}
    H = -\sum_{i=0}^{N-1}\frac{J_{i+\frac{1}{2}}}{2} Z_i Z_{i+1} -\sum_{i=0}^{N-1} \frac{g_i}{2} X_i \equiv H_{Z\!\!\;Z} + H_X,
    \label{eq:H_spin}
\end{align}
where $g_i=\frac{1}{2}(J_{i-1/2}+J_{i+1/2})$. In the absence of spatial deformation, i.e., $J_{i+1/2} \equiv J$, the system sits exactly at the critical point $g=J$, realizing a $c=\frac{1}{2}$ Ising CFT at low energies. 

Inhomogeneous couplings in \eqref{eq:H_spin} that vary slowly on the scale of the lattice spacing yield a deformed TFIM whose long-wavelength description is an \textit{inhomogeneous} CFT~\cite{Dubail_2017_inhcft, 10.21468/SciPostPhys.3.3.019,Gaw_dzki_2018, Moosavi_2021_inh}. As reviewed in the  App.~\ref{sec:deform}, the time evolution of certain observables in the inhomogeneous continuum theory is analytically determined by time-dependent conformal transformations. Therefore, an appropriate choice of couplings in~\eqref{eq:H_spin} yields a Floquet protocol whose intermediate-time dynamics is predicted by CFT. Concretely, we consider a single wavelength deformation~\cite{Han2020-102} (see Fig.~\ref{fig:setup}a)
\begin{equation}
    \label{eq:couplings}
    f(x) = \kappa^0+\kappa^+\cos\!\Big(\frac{2\pi q x}{N}\Big)+\kappa^-\sin\!\Big(\frac{2\pi q x}{N}\Big),
\end{equation}
with wavenumber $q\in\mathbb{N}$ and amplitude parameters $\vec{\kappa}=(\kappa^0,\kappa^+,\kappa^-)\in\mathbb{R}^3$. The lattice couplings are then set by evaluating this envelope at the bond centers, such that $J_{i+1/2} = f(i+1/2)$. 

We design a two-segment drive alternating between the uniform TFIM $H_0$ ($\vec{\kappa}_0=(1,0,0)$) and the deformed TFIM $H_1$, generated by a generic triplet $\vec{\kappa}_1$ via Eq.~\eqref{eq:couplings}. The Floquet unitary is given by
\begin{equation}
    \label{eq:twostepdrive}
    U_F = \mathrm{e}^{-\mathrm{i} H_1 T_1 }\mathrm{e}^{-\mathrm{i} H_0 T_0 }.
\end{equation}
The corresponding continuum theory exhibits a transition between a non-heating and a heating phase, as shown in Fig.~\ref{fig:setup}b. For an initial critical ground state $|\psi_0\rangle$, the time evolution of various observables is constrained by conformal invariance and hence analytically tractable. For example, the energy $E(n) \equiv \langle \psi_0|(U_F^{n})^{\dagger} H_0 U_F^{n}|\psi_0 \rangle$ grows exponentially in the heating phase and oscillates in the non-heating phase. 
Although in the CFT description the energy grows exponentially in the heating phase, the heating dynamics remain controlled by the underlying universality class. 
Moreover, this ``conformal heating'' is characterized by a universal decay of the Loschmidt echo
\begin{equation}
    \mathcal{L}(n) \equiv |\langle\psi_0|U_F^n|\psi_0 \rangle|^2,
\end{equation}
which is explicitly controlled by the central charge $c$ (see App.~\ref{sec:cft_prediction} for details).
Hence, the central charge of any 1D critical model can be extracted from the time evolution of observables such as energy and Loschmidt echo, provided the system is initialized in a critical state. 

\begin{figure*}[t]
    \centering
    \includegraphics[width=0.95\linewidth]{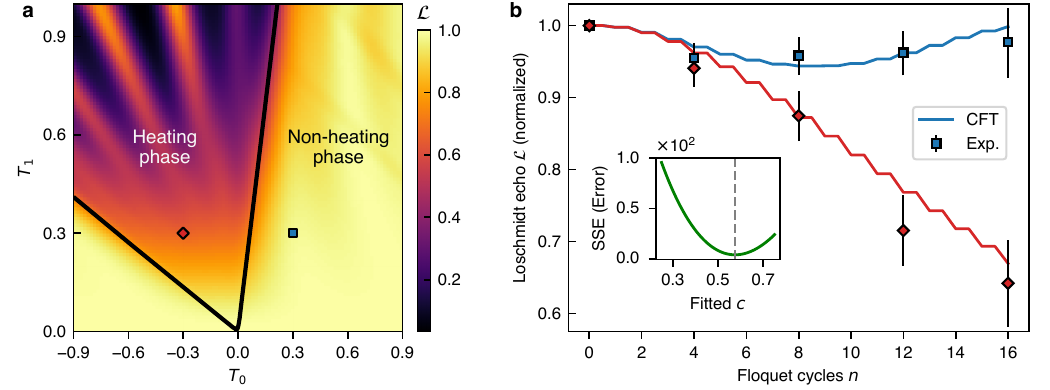}
    \caption{\textbf{Observation of heating and non-heating phases.}
    (a) Phase diagram of the Floquet dynamics, characterized by the Loschmidt echo $\mathcal{L}$ after $n=16$ cycles for $N=16$ spins with spatial deformation parameters $q = 2$, $\vec{\kappa}_1 = (1.0, 1.2, -0.2)$. The black curve denotes the phase boundary predicted by CFT, which separates the heating phase (decaying $\mathcal{L}(n)$) from the non-heating phase (oscillating $\mathcal{L}(n)$). The red diamond and blue square at $T_0 = \mp 0.3$, $T_1 = 0.3$ mark the two representative parameter points used in panel (b). A negative value of $T_0$ corresponds to flipping the sign of the Hamiltonian $H_0$ in the time-evolution operator $\mathrm{e}^{-\mathrm{i}H_0T_0}$.
    (b) Quantum hardware results (Exp.) for the stroboscopic evolution of the normalized Loschmidt echo of the 16-qubit chain, in both the heating phase (red) and the non-heating phase (blue). The initial state corresponds to the critical ground state of the uniform TFIM, prepared by the MERA circuit.
    Each data point is obtained from up to 1000 measurement shots, with error bars representing 95\% confidence intervals estimated using bootstrap resampling. Solid lines show the corresponding CFT predictions. The inset shows the sum of squared errors used to determine the fitted central charge $c$. The measured central charge is $0.575\pm0.036$, which is close to the theory prediction $c=\frac{1}{2}.$}
    \label{fig:echo}
\end{figure*}

\textit{Digital simulation protocol}---The protocol consists of two main stages: (i) preparing the critical ground state of the uniform TFIM $H_0$ as the initial state, and (ii) implementing the Floquet drive via digital Trotterization. 

Preparing critical ground states for quantum many-body systems is a formidable challenge. At the critical point, the system exhibits gapless excitations and algebraically decaying correlations, with quantum entanglement extending across all length scales. These features drastically reduce the efficiency of adiabatic protocols and standard quantum circuits with strictly local gates. To faithfully capture the fundamental nature of these critical states, we employ a hierarchical quantum circuit based on the MERA~\cite{Vidal2007-99,Vidal2008-101}. 

As depicted in Fig.~\ref{fig:setup}c, we prepare the ground state of the critical TFIM $H_0$ at a system size of $N=16$ by applying the unitary $U_\mathrm{MERA}$ to an initial product state of 16 qubits, $|+\rangle^{\otimes 16}$, where $X|+\rangle=|+\rangle$. 
The circuit follows a hierarchical, top-down architecture that builds the state from coarse to fine length scales. It begins at the most macroscopic scale with a single long-range gate (blue) acting on the two top-level representative qubits, \(q_0\) and \(q_8\). Following this, the circuit proceeds through layers of isometries (blue) and entanglers (green). The isometry gates incorporate new degrees of freedom by coupling the previously entangled qubits to the remaining unentangled qubits (e.g., branching $q_0$ into the pair $\{q_0, q_4\}$). Immediately after this branching step, entangler gates are applied between these adjacent branches (e.g., crossing $q_4$ and $q_8$) to capture the entanglement at that specific intermediate scale. 
Repeating this branch-and-entangle procedure layer by layer eventually incorporates all 16 qubits to construct the full lattice state. Such an architecture can efficiently encode long-range correlations and multi-scale entanglement within a circuit of depth $\mathcal{O}(\log N)$.

The hierarchical geometry and isometric constraints of MERA enforce strictly bounded causal cones for local observables, allowing us to classically evaluate the energy and optimize the circuit parameters in polynomial time.
Benchmarking against the exact ground state of the critical TFIM $H_0$ at $N=16$,  the optimized circuit achieves an energy-density error of $7.4\times10^{-4}$ and a state infidelity of $3.1\times10^{-3}$.

With the high-fidelity critical ground state prepared, we implement the Floquet drive defined in Eq.~\eqref{eq:twostepdrive}. Each time-evolution step of duration $\tau$ under the Hamiltonian~\eqref{eq:H_spin} is discretized via a second-order Trotter-Suzuki decomposition, 
\begin{equation} 
    \label{eq:trotter}
    \mathrm{e}^{-\mathrm{i} H \tau} = \mathrm{e}^{-\mathrm{i} H_{ZZ} \tau/2} \mathrm{e}^{-\mathrm{i} H_X \tau} \mathrm{e}^{-\mathrm{i} H_{ZZ} \tau/2} + \mathcal{O}(\tau^3). 
\end{equation} 

\begin{figure*}
    \centering
    \includegraphics[width=0.95\linewidth]{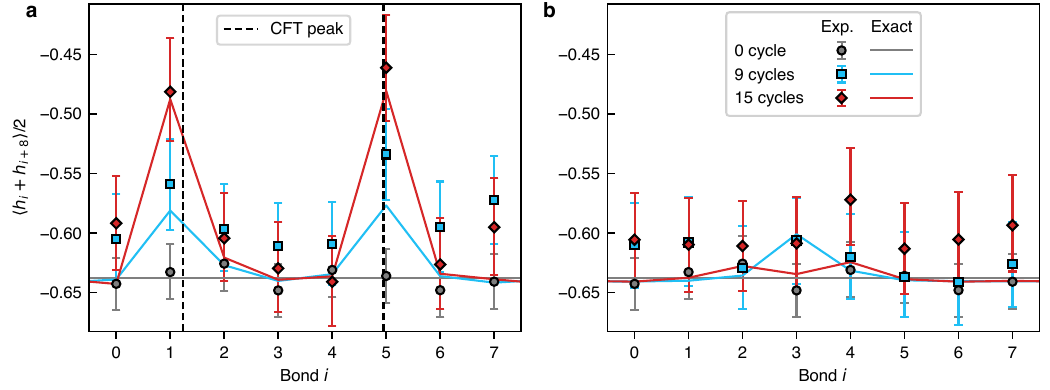}
    \caption{\textbf{Spatial structure of conformal Floquet heating.} (a) Spatial profile of the energy density in the heating phase, demonstrating energy accumulation around the continuous-space locations predicted by CFT (black dashed lines). (b) Corresponding energy density profile in the non-heating phase, which is bounded over Floquet cycles without localized energy accumulation. In both panels, solid lines denote exact numerical simulations of the lattice model (Exact), while markers represent quantum hardware results (Exp.) obtained from up to 1000 measurement shots. Error bars indicate 95\% confidence intervals obtained via bootstrap resampling. To suppress statistical noise, the displayed data points are averaged over $q=2$ sites separated by $N/q$, exploiting the spatial symmetry of the deformation. The parameters are $T_0 = \mp 0.3$, $T_1 = 0.3$, $N = 16$, and deformation triplet $\vec{\kappa}_1 = (1.0, 1.2, -0.4)$.}
    \label{fig:localenergy}
\end{figure*}

To minimize the required circuit depth, we choose short drive durations $|T_0| = T_1 = 0.3$. In this high-frequency driving regime, a single Trotter step suffices for each segment of the Floquet cycle (i.e., setting the Trotter step sizes as $\tau = T_0$ and $\tau = T_1$). Despite this aggressive discretization, the digital evolution remains highly accurate: compared to exact diagonalization simulations, the state infidelity of the non-equilibrium dynamics remains below $3.9\times10^{-2}$ even after $n=16$ Floquet cycles.

\section{Hardware implementation}\label{sec:hardware}

We realize the above protocol on a trapped-ion quantum processor, Quantinuum's Helios~\cite{Ransford2025_11}. Its all-to-all connectivity of atomic platforms provides the crucial flexibility needed to efficiently implement both the non-local gates in MERA and the periodic boundary conditions. The Helios device utilizes $^{137}\text{Ba}^+$ ions, encoding the computational basis in the hyperfine ``clock'' states of the ground $^2S_{1/2}$ electronic manifold as $|0\rangle \equiv |F=1, m_F=0\rangle$ and $|1\rangle \equiv |F=2, m_F=0\rangle$, where $F$ is the total atomic angular momentum and $m_F$ its projection onto the direction of the magnetic field. This specific atomic structure enables the detection of computational subspace leakage via state shelving to the $^2D_{5/2}$ manifold~\cite{Ransford2025_11}. 

Given the substantial depth of our quantum circuits (up to 642 native two-qubit gates in total on 16 qubits), the overall hardware performance is primarily limited by two-qubit gate infidelities (approximately $8 \times 10^{-4}$ for maximally entangling $R_{zz}(\pi/2)$ gates) and idling memory errors (including both linear and quadratic scaling). Leakage contributes significantly to the error budget, accounting for approximately 15--30\% of the $R_{zz}(\pi/2)$ gate error rate and 80\% of the linear memory error rate~\cite{Ransford2025_11}. By discarding measurement shots flagged with leakage, we substantially improve the accuracy of the measured observables.

\section{Loschmidt echo and central-charge extraction}\label{sec:echo}

We characterize the conformal Floquet dynamics by measuring the Loschmidt echo $\mathcal{L}$. Figure~\ref{fig:echo}a shows the exact phase diagram of $\mathcal{L}$ after 16 Floquet cycles as a function of the driving durations $T_0$ and $T_1$, computed for a 16-spin chain with spatial deformation parameters $q = 2$ and $\vec{\kappa}_1 = (1.0, 1.2, -0.2)$. Analytical boundaries derived from CFT (black lines; see  App.~\ref{sec:cft_prediction}) sharply delineate two distinct regimes: a heating phase characterized by monotonic decay of $\mathcal{L}$ over Floquet cycles, and a non-heating phase exhibiting bounded oscillations. To probe these distinct regimes on the Helios quantum processor, we select two representative parameter points ($T_0=\mp 0.3$, $T_1 = 0.3$), marked by the red diamond and blue square, respectively. Note that a negative $T_0$ corresponds to time evolution under $-H_0$.

Figure~\ref{fig:echo}b shows the dynamics of the measured Loschmidt echo $\mathcal{L}(n)$ versus Floquet cycles $n$, displaying the predicted monotonic decay in the heating phase (red) and bounded oscillations in the non-heating phase (blue). To improve data quality, we apply three error-mitigation steps (see App.~\ref{sec:data} for further details). First,  we use embedded dynamical decoupling to suppress coherent phase errors during the digital Floquet drive. Second, we apply a $\mathbb{Z}_2$ filter to post-select measurement shots residing within the $\prod_i^N X_i = +1$ parity sector. Furthermore, to account for residual decoherence, we implement reference circuits that share the exact same structure, but replace the deformed driving Hamiltonian with the uniform one. In the latter case, the prepared ground state evolves entirely under its parent Hamiltonian and remains unchanged up to a global phase. The Loschmidt echo should therefore be unity in the ideal case, and we attribute any observed decay to hardware noise, which is empirically fit by $0.991(3)\exp(-0.0068(6)n)$. Normalizing the measured Loschmidt echo by this empirical fit compensates for the background decoherence. The resulting hardware data (markers) are in quantitative agreement with the CFT results (solid lines) for both the heating and non-heating regimes.

Having verified the conformal character of the observed dynamics, we proceed to extract the central charge, $c$, a fundamental parameter that characterizes the universality class of the underlying lattice model. As detailed in the App.~\ref{sec:cft_prediction}, we obtain a concise closed-form expression for the Loschmidt echo:
\begin{equation}\label{eq:echo}
    \mathcal{L}(n) = |\alpha_n|^{-\frac{q^2-1}{3q}\,c}, 
\end{equation}
where $\alpha_n$ is a matrix element of the M\"obius transformation governing the Floquet evolution (see App.~\ref{sec:cft_prediction}) and depends exclusively on the driving parameters ($T_0$, $T_1$, $\vec{\kappa}_1$, $q$). While the bounded oscillations in the non-heating phase exhibit a weak pattern that prevents robust parameter estimation, the pronounced decay in the heating phase provides a strong signal for reliable fitting. By minimizing the sum of squared errors (SSE) between our measured data in the heating phase and the exact formula~\eqref{eq:echo}, we extract a central charge of $c = 0.575 \pm 0.036$ (Fig.~\ref{fig:echo}b, inset), which is close to the theoretical prediction of $c = 1/2$ for the critical Ising class.

\section{Spatial structure of heating}\label{sec:energy}

Unlike generic driven many-body quantum systems, which typically evolve toward a featureless state, the driven CFT in the heating phase develops a nontrivial spatial structure. Specifically, the energy density and entanglement entropy develop $2q$ sharp peaks, which correspond to sinks of gapless quasiparticles~\cite{Fan2020-10}. Thus, the system only \textit{locally} heats up, while the remainder effectively cools down below the ground-state energy density, giving rise to a dynamical Casimir effect~\cite{PhysRevLett.73.1931, Fan2020-10, Martin_2019_casimir}. In contrast, the system fails to locally accumulate energy in the non-heating phase. Here, the energy density oscillates periodically, and the state exhibits periodic finite-time revivals to the initial ground state, where the period diverges at the heating phase transition.

Although CFT is defined in continuous spacetime, where injected energy in the heating phase grows exponentially and without bound, it significantly constrains the physics of discrete lattice models, even at small system sizes.
In our demonstration, we implement the driven Ising chain with $\vec{\kappa}_1 = (1.0, 1.2, -0.4)$. 
As shown in Fig.~\ref{fig:localenergy}a, the measured local energy density $h_i = -\frac{1}{2}Z_iZ_{i+1}-\frac{1}{4}(X_i+X_{i+1})$ in the heating phase strongly accumulates on the lattice bonds closest to the continuous-space locations predicted by CFT.
In contrast, the non-heating phase (Fig.~\ref{fig:localenergy}b) does not feature such localized energy accumulation. While hardware noise and statistical uncertainties mask the predicted oscillations of the energy density, our data clearly demonstrate that the energy remains bounded. 
Here, we present the raw data directly, and the embedded dynamical decoupling during the digital Floquet drive is the only error mitigation strategy we employ.

\section{Discussion and Outlook}\label{sec:outlook}

In this work, we have demonstrated the emergence of conformal Floquet dynamics in a TFIM chain on a digital trapped-ion quantum processor. This emergent behavior is generic and not unique to the TFIM. Numerical simulations in App.~\ref{sec:generic} show agreement with CFT predictions across different universality classes, including the non-integrable interacting Ising chain ($c = 1/2$) and the integrable XX chain ($c = 1$). Although our current hardware and numerical benchmarks are carried out on a 16-qubit lattice, the discrete lattice regularization approaches the continuum limit of CFT as the system size $N$ grows. The underlying physics is therefore robust against finite-size effects and is expected to manifest even more cleanly on larger scales.

Our protocol is furthermore scalable across all three constituent operations: state preparation, Floquet evolution, and measurement. First, the classical computational complexity of optimizing the MERA state preparation circuit is bounded by $\mathcal{O}(N)$, owing to the strictly bounded causal cones of the MERA architecture. Second, the digital Floquet driving circuit is straightforward to implement, requiring a total driving depth that scales only linearly with $N$ to allow gapless excitations to propagate across the system~\cite{Fan2020-10}. Finally, extracting universal data via our protocol is also scalable; whereas determining the central charge via subsystem entanglement entropy generically incurs an exponential overhead in terms of measurement bases, we extract $c$ directly from the Loschmidt echo which requires a single measurement basis. 
Alternatively, for sufficiently large systems, $c$ can be extracted from the dynamics of the total energy, a local observable that can be estimated even more robustly.
 
Our protocol also offers a physics-driven verification framework for quantum hardware in regimes that lie beyond classical simulability. As digital quantum processors scale past 50--100 qubits, standard numerical cross-checking rapidly becomes impractical. However, because our protocol yields closed-form CFT predictions for the non-equilibrium dynamics, it establishes a reliable benchmarking tool for large-scale hardware: one that verifies whether the intermediate-time driven dynamics conform to the CFT predictions without requiring classical lattice simulations.

More broadly, as demonstrated here for a critical initial state and in Ref.~\cite{Chertkov2022-18} for a non-critical one, the synergy between efficient tensor-network state preparation and digital quantum evolution offers a generic framework for exploring non-equilibrium dynamics at large scale. Beyond the exact setting realized here, this framework naturally points to several regimes where CFT or other analytical theories no longer provide direct predictions.
(i) \textit{Driving non-conformal states:} preparing and driving critical states without exact CFT descriptions, such as Lifshitz-type criticality~\cite{Hornreich1975-35,Saito2024-132}, would generate new dynamical data rather than confirm known predictions.
(ii) \textit{Non-conformal driving:} periodically driving critical ground states with off-critical relevant operators. Although such drives are generally expected to induce nonuniversal heating toward a featureless state, whether universal signatures persist at early times remains an open question.
(iii) \textit{Probing the emergent bulk dynamics:} in holographic tensor-network constructions, the renormalization direction plays the role of an emergent bulk dimension. Sahay \textit{et al.}~\cite{Sahay2025-15} recently showed that a MERA representation for a critical Ising chain yields gravity-like attractive interactions between bulk excitations. This is a static result, and whether these interactions give rise to gravity-like dynamics, such as infall or collision, remains an open question. Our work has already prepared the same critical state, except for flipping a few initial qubits to introduce bulk excitations; driving such states in real time offers a route toward probing emergent gravitational dynamics on quantum hardware.

\section*{Acknowledgments}

We thank Biao Lian, Sara Murciano and Peter Zoller for fruitful discussions. We acknowledge the use of Quantinuum services, and thank the entire Quantinuum team for their contributions toward the successful operation of the Helios processor.
This research used resources from the Oak Ridge Leadership Computing Facility at the Oak Ridge National Laboratory, which is supported by the Office of Science of the U.S. Department of Energy under Contract No. DE-AC05-00OR22725.

\textit{Author contributions}---L.M., B.L., and Q.M. conceptualized the work. L.M. and B.L. led the theoretical efforts, while L.M. and Q.M. led the numerical simulations. Q.M. designed and executed the hardware implementation and analyzed the data. All authors contributed to the manuscript.

\textit{Data availability}---All data needed to evaluate the conclusions in this paper are present in the main text or the appendices.
Numerical codes supporting the findings are publicly accessible at~\hyperlink{https://github.com/LianghongMo/Driven-CFT}{https://github.com/LianghongMo/Driven-CFT}.

\appendix

\begin{figure}
    \centering
    \includegraphics[width=0.9\linewidth]{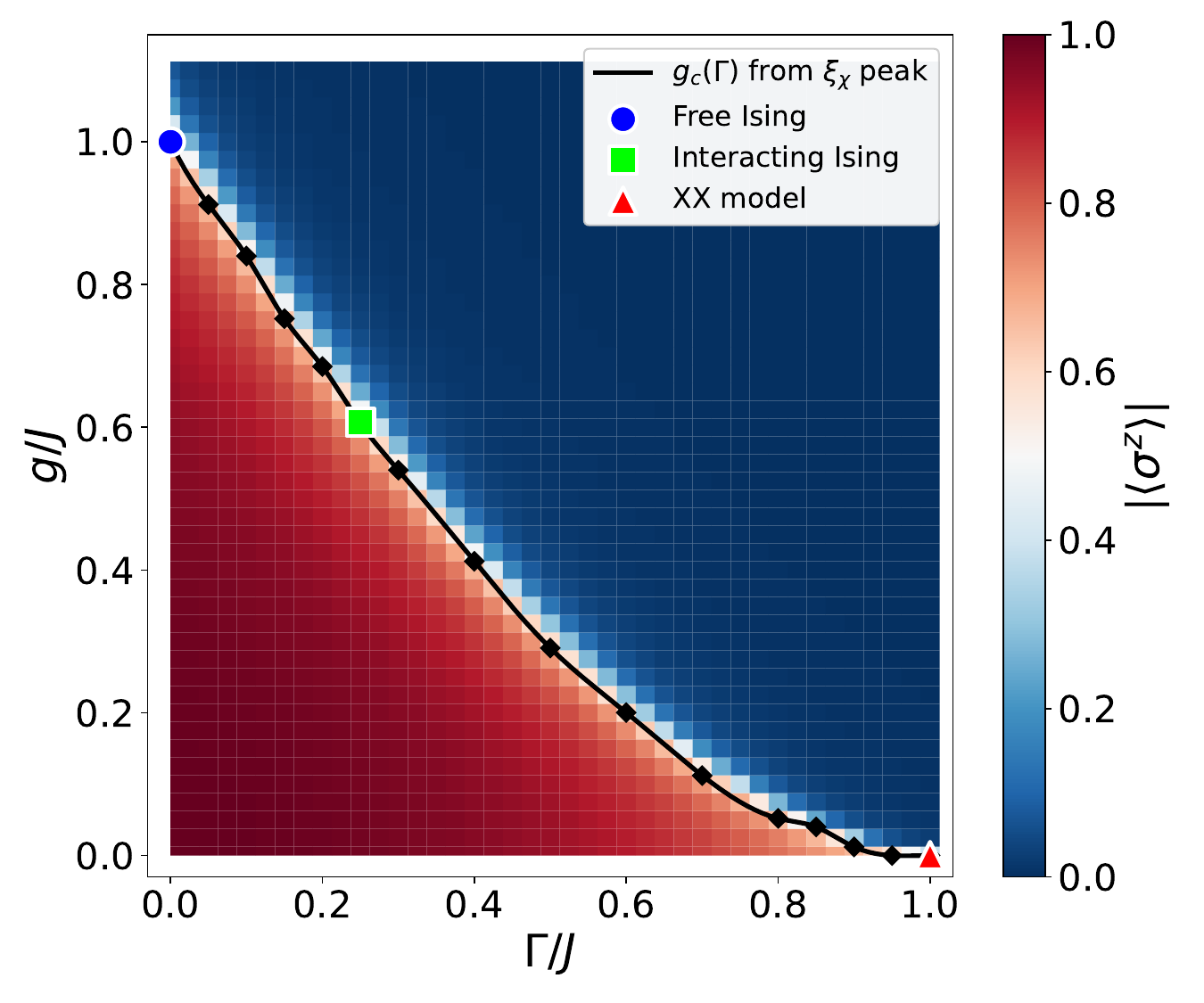}
    \caption{\textbf{Phase diagram of the perturbed transverse-field Ising
    chain~\eqref{eq:TFIMnon-integrable}}. The colour map shows the order parameter
    $|\langle\sigma^z\rangle|$ across the $(\Gamma, g)$ plane; the black
    curve is the critical boundary $g_c(\Gamma)$ extracted from the
    $\xi_\chi$ peak. The three
    highlighted markers are benchmark points with known central charge: the
    free Ising critical point $(\Gamma, g) = (0, 1)$ ($c = 1/2$),
    the ``interacting'' Ising critical point $(\Gamma, g) = (0.25, 0.607)$
    ($c = 1/2$), and the XX model $(\Gamma, g) = (1, 0)$ ($c = 1$).}
    \label{Fig:orderparmphasediagram}
\end{figure}

\begin{figure*}
    \centering
    \includegraphics[width=0.95\linewidth]{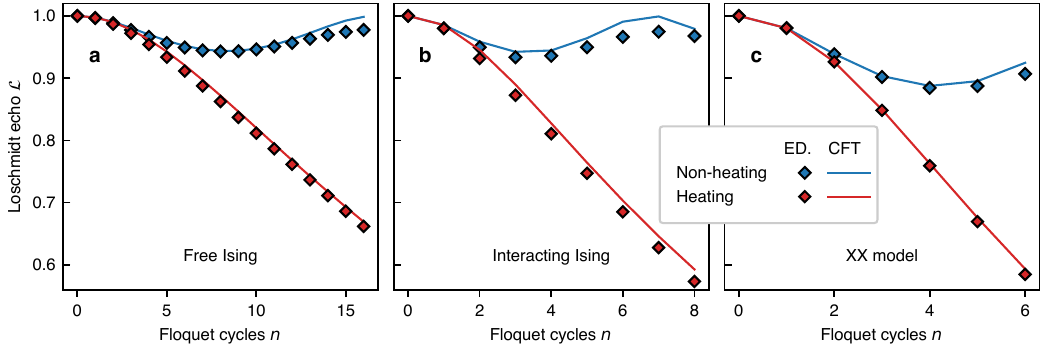}
    \caption{\textbf{Comparison between exact diagonalization and CFT predictions for different critical lattice models obtained from \eqref{eq:TFIMnon-integrable}.} (a) Free TFIM, $(J, g, \Gamma) = (1/2, 1/2, 0)$. (b) Non-integrable interacting Ising model, $(J, g, \Gamma) = (1, 0.6066, 0.25)$. (c)  XX model, $(J, g, \Gamma) = (1/2, 0, 1/2)$. Markers denote exact diagonalization calculations with size $L = 16$ and periodic boundary conditions.
    Solid lines are the analytical CFT predictions from Eq.~\eqref{eq:echo}.
    Here we use deformation $\vec\kappa_1 = (1.0, 1.2, -0.2)$, $q = 2$, with $T_0=-0.3$, $T_1=0.3$ for heating phase and $T_0=0.3$, $T_1=0.3$ for non-heating phase. 
    Across these three microscopically different models, numerical data at $L=16$ agree quantitatively with the CFT prediction in both phases, demonstrating that the protocol is universal and depends only on the underlying CFT data $c$, both for integrable and non-integrable models.}
    \label{Fig:fidelity_3panels}
\end{figure*}

\begin{figure*}
    \centering
    \includegraphics[width=0.95\linewidth]{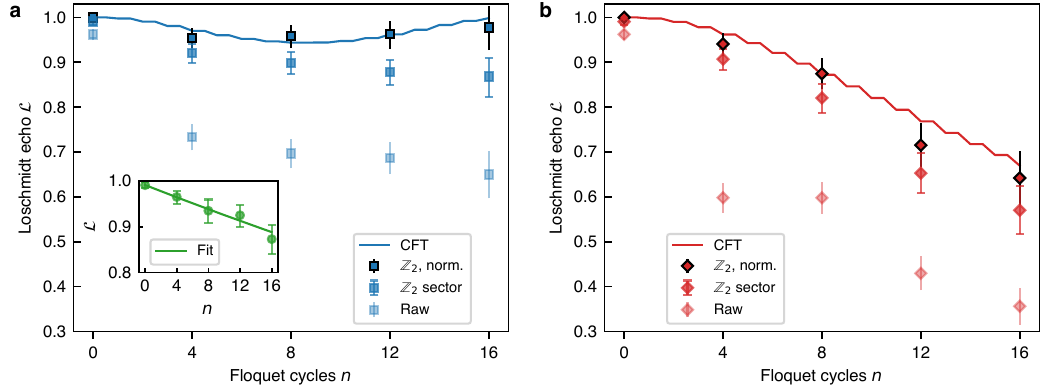}
    \caption{\textbf{Error mitigation and data processing for the measured Loschmidt echo.} (a) Non-heating dynamics. The blue solid line is the CFT prediction. The hardware data are shown at three levels of processing: raw measurement outcomes (light blue), data post-selected into the positive parity $\mathbb{Z}_2$ sector (blue) and the final data after dividing out the reference decay (dark blue). Inset: reference echo under the uniform Hamiltonian, whose decay is fitted and used for the normalization  procedure. (b) Heating dynamics with the same analysis procedure as the non-heating case. 
    In all plots, error bars denote 95\% confidence intervals obtained via bootstrap resampling.}
    \label{fig:dataProcess}
\end{figure*}

\section{Numerical results on other driven critical models}\label{sec:generic}

In this section, we demonstrate the universality of the Loschmidt echo time evolution in both heating and non-heating phases by numerically studying different critical lattice models.
Concretely, we numerically benchmark the CFT predictions on the non-integrable Ising chain
\begin{align}
    H = -\sum_i \bigl[J\,Z_iZ_{i+1} + g \,X_i + \Gamma\,X_iX_{i+1}\bigr],
    \label{eq:TFIMnon-integrable}
\end{align}
where the spatial deformation can be applied to each term, similar to Eq. \eqref{eq:H_spin}. 
By tuning the parameters $(J,g,\Gamma)$, this Hamiltonian interpolates among three distinct models with $\mathbb{Z}_2$ symmetry: the free TFIM at $(J,g,\Gamma)= (1/2,1/2,0)$, studied in the main text; a non-integrable Ising model at $(J,g,\Gamma)= (1,0.6066, 0.25)$, sharing the same CFT data with $c=1/2$; and the XX model at $(J,g,\Gamma)= (0,0,1/2)$, equivalent to a complex free Fermion model with $c=1$.

We compute the phase diagram at zero temperature using infinite Density Matrix Renormalization Group (iDMRG) on a two-site unit cell following Ref.~\cite{cole2017entanglement}. The critical line $g_c(\Gamma)$ is derived from the peak of the effective correlation length $\xi_\chi(g) \equiv -1/\ln|\lambda_2|$, where $\lambda_2$ is the second-largest eigenvalue of the iMPS transfer matrix. $\xi_\chi$ saturates at finite $\chi$ but diverges as $\chi\!\to\!\infty$ at criticality. For each $\Gamma$, we perform a multi-$\chi$ scan ($\chi \in \{16, 24, 32, 48\}$) over a fine $g$ grid and take the parabolic peak position at the largest $\chi$ as $g_c$. To visualize the order parameter $|\langle\sigma^z\rangle|$ across the phase diagram (Fig.~\ref{Fig:orderparmphasediagram}), we run a separate uniform $(\Gamma, g)$ scan at $\chi = 32$ with a small $\mathbb{Z}_2$-breaking pinning field $\epsilon_z = 10^{-3}$ that selects one of the two symmetry-broken states in the ferromagnetic phase. The three benchmark models all sit on the critical line.

We then evolve each of the three critical ground states under Floquet drives, with the deformation parameters $\vec{\kappa}_1 = (1.0, 1.2, -0.2)$. For the heating phase, we use $T_0 =-0.3,T_1=0.3$ and for the non-heating phase, $T_0=T_1=0.3$.   Note that this should be rescaled as $T_{\mathrm{CFT}}=vT_{\mathrm{lattice}}$ by the non-universal velocity, which can be extracted from the finite-size  spectrum~\cite{xavier2010entanglement,berdanier2017floquet}.

Figure~\ref{Fig:fidelity_3panels} compares the exactly computed evolution for the Loschmidt echo at $L=16$ with the CFT prediction  in Eq.~\eqref{eq:echo} (see App.~\ref{sec:cft_prediction} for details). In the heating phase, the echo decays exponentially, while in the non-heating phase, it oscillates with a bounded amplitude. Both behaviors are captured by the CFT prediction within a few percent error over the entire range of cycles for all three models, despite the different microscopic Hamiltonians.
This confirms that our non-equilibrium protocol for extracting $c$ is universal: critical models in the same universality class follow the same Loschmidt echo, as predicted by CFT. This includes strongly correlated, non-integrable cases where no analytical lattice solution exists.

\section{Error mitigation and data processing}\label{sec:data}

In this section, we describe the main steps we have employed to produce Fig.~\ref{fig:echo} in the main text. It includes three error mitigation steps: embedded dynamical decoupling, $\mathbb{Z}_2$ parity post-selection, and reference-circuit normalization, suppressing both coherent phase noise and decoherence in the measured Loschmidt echo.

(i) \textit{Embedded dynamical decoupling}: A dominant source of error in trapped-ion processors, such as Quantinuum's Helios, is coherent phase noise. This arises from qubit frequency offsets driven by residual magnetic fields and calibration imperfections~\cite{Ransford2025_11}. Standard Dynamical Decoupling (DD)~\cite{Viola_1999} mitigates these slowly varying offsets by inserting echo pulses during qubit idling windows to effectively flip the sign of the offset. In this work, however, we implement an embedded DD approach. Owing to the highly efficient logarithmic depth of the MERA circuit, the leading circuit errors reside in the subsequent Trotterized time-evolution sequence, which consists of alternating layers of two-qubit $ZZ$ gates and single-qubit $R_x$ rotations across all qubits. To natively embed the DD within this sequence, we shift the rotation angles of all $R_x(\theta)$ gates to $R_x(\theta \pm \pi)$, while simultaneously inverting the sign of the adjacent $ZZ$ gates to maintain the theoretical invariance of the circuit. Because the target rotation angles are intrinsically small, this embedded DD technique successfully cancels the leading-order accumulated dephasing without requiring additional gate overhead.

(ii) \textit{Post-selection and normalization}: On hardware, the Loschmidt echo $\mathcal{L}$ is evaluated by measuring the overlap between the initial product state $|+\rangle^{\otimes N}$ and the evolved state $U_{\text{MERA}}^\dagger U_F^n U_{\text{MERA}}|+\rangle^{\otimes N}$. Since both the MERA state preparation circuit $U_{\text{MERA}}$ and the Floquet time-evolution circuit $U_F^n$ preserve $\mathbb{Z}_2$ symmetry, we post-select the raw data by retaining only the measurements shots that fall within the positive parity sector ($\prod_i^N X_i = +1$)~\cite{Bonet_Monroig_2018}. As shown in Fig.~\ref{fig:dataProcess}, this $\mathbb{Z}_2$ parity filtering yields a significant improvement over the raw data in both the heating and non-heating phases.

However, this filter cannot mitigate symmetry-preserving errors, such as correlated $ZZ$ noise. To address this, as discussed in the main text, we utilize reference circuits to estimate the residual noise. By replacing the deformed Hamiltonian with a uniform Hamiltonian, we fit the decay of the reference $\mathcal{L}$ to an empirical decay function. Finally, by rescaling the $\mathbb{Z}_2$-sector data against this empirical function, we obtain a normalized $\mathcal{L}$ that demonstrates good agreement with theoretical predictions.

\section{Deformed Hamiltonians and M\"obius evolution}\label{sec:deform}
We begin by briefly reviewing the continuous-time dynamics of a one‑dimensional conformal field theory (CFT) whose stress tensor is spatially deformed. A generic deformed Hamiltonian reads
\begin{equation}\label{Eq:H}
H = \frac{1}{2\pi}\int_0^L \left[f(x)\,T(x)+\bar f(x)\,\bar T(x)\right] \, \mathrm{d}x,
\end{equation}
where $T(x)$ and $\bar T(x)$ are the chiral and antichiral components of the stress tensor. In order to describe a drive that can effectively be realized in a lattice model, we set $\bar f(x)=f(x)$. For simplicity, we restrict this spatial profile to a single harmonic sector~\cite{Han2020-102}, taking the explicit form
\begin{align}
f(x)=\kappa^0+\kappa^+\cos\!\left(\frac{2\pi q x}{L}\right)+\kappa^-\sin\!\left(\frac{2\pi q x}{L}\right),
\qquad q\in \mathbb{N},
\label{Eq:Hamiltonian}
\end{align}
with \(\vec{\kappa} = (\kappa^0,\kappa^+,\kappa^-) \in \mathbb{R}^3\). 
Using the mode expansion $T(x)=\frac{2\pi}{L}\sum_n L_n \mathrm{e}^{-2\pi \mathrm{i} n x/L}$,
the deformed Hamiltonian \eqref{Eq:H} with profile \eqref{Eq:Hamiltonian} at mode $q$ becomes
\begin{eqnarray}
H_q = &\frac{2\pi}{L}&\!\left[
   \kappa^0 L_0
   +\frac{\kappa^+}{2}(L_q+L_{-q})
   +\frac{\kappa^-}{2\mathrm{i}}(L_q-L_{-q})
\right]\nonumber\\
  &+&\text{antichiral},
\end{eqnarray}
which closes in the $\mathfrak{su}(1,1)$ algebra generated by $\{L_0,\, L_q,\, L_{-q}\}$ (and their antichiral counterparts).

The key consequence is that the Heisenberg time evolution of primary fields under any fixed Hamiltonian in this family for a duration $t$ can be represented exactly by a M\"obius transformation on the complex coordinates. This geometric action is encoded in an SU$(1,1)$ matrix, which takes the general form
\begin{align}
G = \begin{pmatrix} a &b\\ b^\ast & a^\ast \end{pmatrix}, \qquad |a|^2-|b|^2=1.
\end{align}
For the chiral sector, its M\"obius matrix $G(t)$ has elements
\begin{align}
\label{eq:coefmobiss}
a(t) &= -\cos\!\left(s \frac{\pi q t}{L}\right) - \mathrm{i}\frac{\kappa^0}{s}\sin\!\left(s \frac{\pi q t}{L}\right), \\
b(t) &= -\mathrm{i}\frac{\kappa^+ + \mathrm{i}\kappa^-}{s}\sin\!\left(s \frac{\pi q t}{L}\right),
\label{eq:coefmobiss2}
\end{align}
where $s = \sqrt{(\kappa^0)^2 - (\kappa^+)^2 - (\kappa^-)^2}$~\footnote{With the identities $\cos(\mathrm{i}x)=\cosh(x)$ and $\sin(\mathrm{i}x)=\mathrm{i}\sinh(x)$, these unified trigonometric expressions naturally accommodate the hyperbolic case when $s^2 < 0$.}. Notice that the M\"obius matrix only depends on the dimensionless ratio $t/L$, hence any physical observables are only functions of this ratio.
The antichiral matrix $\bar{G}(t)$ is simply related to the chiral sector by $\bar{a}(t)=a(t)$ and $\bar{b}(t)=-b^\ast(t)$.  

The continuous-time representation can be extended to describe generic piecewise-constant driven dynamics. Consider a Floquet dynamics with $U_0(T_0)=\exp(-\mathrm{i}H_0T_0)$ and $U_1(T_1)=\exp(-\mathrm{i}H_1T_1)$, whose corresponding M\"obius matrices are $G_0(T_0)$ and $G_1(T_1)$, respectively.  The net chiral evolution $U_F(T)=U_1(T_1)U_0(T_0)$ after one cycle of duration $T=T_0+T_1$ is then represented by the product of the individual M\"obius matrices,
\begin{align}
\Pi(T) = G_0(T_0) G_1(T_{1}).
\end{align}
The net antichiral evolution is similarly represented by the product matrix $\bar{\Pi}(T) = \bar{G}_0(T_0) \bar{G}_1(T_1)$.
Since $\Pi(T)$ encodes a single cycle of the Floquet evolution (for instance the 2-step drive discussed in the main text), the long-time dynamics of the system is strictly classified by the trace of the one-cycle matrix, $\operatorname{Tr}(\Pi(T))$. The Floquet phases fall into three distinct regimes~\cite{wen2018_05, Fan2020-10}: 
\begin{itemize}
    \item Non-heating phase ($|\operatorname{Tr} \Pi(T)| < 2$): The M\"obius transformation is elliptic, leading to bounded, oscillatory dynamics.
    \item Heating phase ($|\operatorname{Tr} \Pi(T)| > 2$): The transformation is hyperbolic, acting as a source and sink for the coordinate flow. This drives an exponential divergence of the M\"obius matrix elements.
    \item Critical point ($|\operatorname{Tr} \Pi(T)| = 2$): The transformation becomes parabolic, marking the phase boundary where the dynamics exhibit algebraic growth.
\end{itemize}
In the next section, we will describe how to relate the coefficients of the one-cycle transformation to observables in the CFT, such as energy density and Loschmidt echo.

\section{Time evolution of observables}\label{sec:cft_prediction}
In this section, our goal is to derive formulas for the time evolution of the Loschmidt echo and energy density for the 2-step drive introduced above.
Our key ingredient will be that, since both $H_0$ and $H_1$ are combinations of $\mathfrak{sl}(2)$ generators, their action preserves a family of coherent states, defined as
\begin{equation}
|\eta\rangle = \mathrm{e}^{\xi L_{-q}-\xi^{\ast} L_{q}}|0\rangle,
\end{equation}
where $\xi\in\mathbb{C}$, $q\in\mathbb{N}_{>0}$, and $|0\rangle$ is the ground state of a (1+1)d chiral CFT on a cylinder of circumference $L$. Each coherent state is uniquely labeled by a variable $\eta=\frac{\xi}{|\xi|}\tanh(q|\xi|)$ in the unit disk; in particular, the ground state corresponds to $\eta=0$, and a state with infinite energy corresponds to $|\eta|\rightarrow1$.
We stress that, in general, the antichiral part of the theory also needs to be included. 
The action of  $n$-cycle Floquet unitary operator on a coherent state is~\cite{lapierre2025driven},
\begin{equation}
U^n_F(T)|\eta\rangle = \mathrm{e}^{\mathrm{i}\phi}|\eta_n\rangle, \quad \eta_n = \frac{\alpha_n\eta+\beta_n}{\beta^{\ast}_n\eta+ \alpha^{\ast}_n},
\label{eq:transformationlaw}
\end{equation}
where $\mathrm{e}^{\mathrm{i}\phi}$ is an irrelevant $\mathrm{U}(1)$ phase, and the coefficients of the time-dependent Möbius transformation are determined by
\begin{align}
    \underbrace{\Pi(T)\,\Pi(T)\cdots\Pi(T)}_{n\text{ times}} =: \begin{pmatrix} \alpha_n & \beta_n\\ \beta_n^\ast & \alpha_n^\ast \end{pmatrix}.
\end{align}
Therefore, the time evolution of any correlator can be obtained from the general properties of such coherent states. The energy density of a coherent state is given as~\cite{Caputa_2023, Liska_2023}
\begin{equation}
    \langle \eta | T(x)|\eta\rangle=\frac{\frac{c}{24}(q^2-1)z^{2(q-1)}(1-|\eta|^2)^2}{(z^q-\eta)^2(1-z^q \eta^*)^2}-\frac{c}{24}(q^2-1)\frac{1}{z^2},
\end{equation}
with $z=\mathrm{e}^{2\pi \mathrm{i} x/L}$ being the spatial coordinate on the cylinder. Therefore, any time evolution of the coherent state, which amounts to replacing $\eta$ by $\eta_n$, is given by
\begin{eqnarray}
\mkern-45mu\langle T(x)\rangle_n \mkern-1mu &\equiv& \mkern-1mu \langle \eta_n| T(x)|\eta_n\rangle \nonumber\\
  &=&\mkern-1mu\frac{c(q^2-1)}{24}\mkern-3mu\left(\mkern-2mu\frac{z^{2(q-1)}(1-|\eta_n|^2)^2}{(z^q-\eta_n)^2(1-z^q \eta^*_n)^2}-\frac{1}{z^2}\mkern-4mu\right)\mkern-3mu.
\end{eqnarray}
Performing a similar computation for the antichiral part of the CFT and using the fact that the initial state is the ground state yields
\begin{subequations}
\label{Eq:EnergyDensities}
\begin{align}
    \frac{1}{2\pi}\langle T(x) \rangle_n &= -\frac{\pi c}{12L^2}q^2 + \frac{\pi c}{12L^2}(q^2-1) \frac{1}{\left|\alpha_n\,\mathrm{e}^{\mathrm{i}\frac{2\pi q x}{L}} + \beta_n\right|^4}, \label{Eq:ChiralDensity} \\
    \frac{1}{2\pi}\langle \bar{T}(x) \rangle_n &= -\frac{\pi c}{12L^2}q^2 + \frac{\pi c}{12L^2}(q^2-1)\frac{1}{\left|\bar{\alpha}_n\,\mathrm{e}^{-\mathrm{i}\frac{2\pi q x}{L}}+\bar{\beta}_n\right|^4},
    \label{Eq:AntichiralDensity}
\end{align}
\end{subequations}
such that the full spatial energy density is
\begin{align}
    \varepsilon^{(n)}_{\mathrm{CFT}}(x)=\frac{1}{2\pi}\!\left(\langle T(x) \rangle_n+ \langle \bar T(x)\rangle_n\right).
\end{align}
Integrating this density over the system length $L$ yields the exact total energy,
\begin{equation}
    E_{\mathrm{CFT}}(n) = -\frac{\pi c}{6L}q^2 + \frac{\pi c}{12L}(q^2-1) \big(|\alpha_n|^2+|\beta_n|^2 + |\bar{\alpha}_n|^2+|\bar{\beta}_n|^2\big).
\end{equation}
Consequently, the total energy oscillates in the non-heating phase and becomes exponentially unbounded in the heating phase. Geometrically, the heating originates from the emergence of a fixed point in the Floquet map at the boundary of the disk in the heating phase, which implies that $\lim_{n\rightarrow\infty}|\eta_n|=1$.

Finally, we compute the Loschmidt echo, defined as the overlap between the time-evolved state and the initial state, $\mathcal{L}_{\rm CFT}(n) \equiv \big|\langle \psi_0|\psi_n\rangle\big|^2,|\psi_n\rangle = U^n_F(T)|\psi_0\rangle$. Given that any time-evolved coherent state remains coherent~\cite{lapierre2026nonequilibriumprobesquantumgeometry}, it suffices to invoke the inner product between coherent states, i.e.,
\begin{align}\label{eq:inner}
 \langle \eta_1|\eta_2\rangle
= \left(\frac{1-\eta_1^*\,\eta_2}
        {\sqrt{(1-|\eta_1|^2)(1-|\eta_2|^2)}}\right)^{\!-\frac{(q^2-1)c}{12q}}.
\end{align}
Let us now assume that the initial state is the ground state of the uniform Hamiltonian. Therefore, using Eq.~\eqref{eq:transformationlaw}, we have $ \eta_n = \frac{\beta_n}{\alpha^{\ast}_n}$. Plugging this back into the inner product formula \eqref{eq:inner}, and combining both chiral and antichiral sectors leads to 
\begin{eqnarray}\label{Eq:loschmidtecho}
    \mathcal{L}_{\rm CFT}(n) &=& \mathcal{L}_{\rm CFT,chiral}(n) \, \mathcal{L}_{\rm CFT,antichiral}(n) \nonumber\\ 
    &=& |\alpha_n|^{\frac{-(q^2-1)c}{6q}} \cdot |\bar{\alpha}_n|^{\frac{-(q^2-1)c}{6q}} \nonumber\\
    &=&  |\alpha_n|^{-\frac{q^2-1}{3q}\,c}.
\end{eqnarray}
This formula leads to oscillations between 1 and a finite value in the non-heating phase, and to monotonic decay towards 0 in the heating phase. In particular, because $\alpha_n$ is uniquely determined by the driving Hamiltonians, the only theory-dependent piece in the Loschmidt echo is the central charge $c$. This feature enables the central-charge extraction demonstrated in the main text.

\bibliography{drivenCFT_ref}

\end{document}